\documentclass[conference]{IEEEtran}  

\title{A Stochastic Geometry Framework for LOS/NLOS Propagation in Dense Small Cell Networks}
\author{Carlo Galiotto$^1$,~~Nuno K. Pratas$^2$,~~Nicola Marchetti$^1$,~~Linda Doyle$^1$\\
1) CTVR, Trinity College, Dublin, Ireland \\
2) Department of Electronic Systems,
Aalborg University (AAU), Denmark \\ 1) \{galiotc,~marchetn,~Linda.Doyle\}@tcd.ie,~~~~2) nup@es.aau.dk.
}

\usepackage[dvips]{graphicx}
\usepackage[T1]{fontenc}
\usepackage{amsmath}
\usepackage{amssymb}
\usepackage{amsthm}
\usepackage{amssymb}
\usepackage{amsfonts}
\usepackage{cite}
\usepackage[nolist]{acronym}
\usepackage{microtype}
\usepackage{tabularx}
\usepackage{multirow}
\usepackage{latexsym}
\usepackage{eurosym}
\usepackage{here}
\usepackage{soul} 		
\usepackage{epsfig}
\usepackage{color}      
\usepackage{algorithmic}
\usepackage{subfigure,stfloats}         
\usepackage{float}                    
\usepackage{lettrine}                

\begin{document}

\bstctlcite{IEEEexample:BSTcontrol}
\maketitle
\begin{abstract}

The need to carry out analytical studies of wireless systems often motivates the usage of simplified models which, despite their tractability, can easily lead to an overestimation of the achievable performance. In the case of dense small cells networks, the standard single slope path-loss model has been shown to provide interesting, but supposedly too optimistic, properties such as the invariance of the outage/coverage probability and of the spectral efficiency to the base station density. 

This paper seeks to explore the performance of dense small cells networks when a more accurate path-loss model is taken into account. We first propose a stochastic geometry based framework for small cell networks where the signal propagation accounts for both the Line-of-Sight (LOS) and Non-Line-Of-Sight (NLOS) components, such as the model provided by the 3GPP for evaluation of pico-cells in Heterogeneous Networks. We then study the performance of these networks and we show the dependency of some metrics such as the outage/coverage probability, the spectral efficiency and Area Spectral Efficiency (ASE) on the base station density and on the LOS likelihood of the propagation environment. 
Specifically, we show that, with LOS/NLOS propagation, dense networks still achieve large ASE gain but, at the same time, suffer from high outage probability. 

\end{abstract}




\vspace{-1mm}
\section{Introduction} 
\label{sect:intro}

The evaluation of wireless communications systems commonly resorts to the use of simplified channel models, with the purpose of simplifying the analytical formulation of the associated system model. While simplified channel models are more tractable, they can easily lead to inaccurate results and consequent wrong conclusions.

A concrete example is the case of cell-splitting in cellular networks, where the network performance has been assessed using a single slope path loss model. By considering this model, it has been shown that the Area Spectral Efficiency (ASE) increases linearly with the Base Station (BS) deployment density, whereas the outage probability is independent of the BS density~\cite{Andrews2011}. Moreover, the energy efficiency turns out to be a monotonic increasing function of the cell density \cite{Galiotto2014}. Nonetheless, when the assumption of single slope path-loss is dropped and different path-loss models are used, those aforementioned properties do no longer hold \cite{Galiotto2014,Ling2012,Zhang2014}. 

Although the results obtained with single slope path loss can be considered reliable if the BS density is within a limited range of values, the same might not be true if one needs to extend this result to a wider BS density range. Therefore, as the trend of future networks is shifting towards ultra dense base station deployment~\cite{Bhushan2014}, it becomes important to re-evaluate the performance of these network deployments using more accurate channel models.

\subsection{Related Work} 
\label{sub:related_work}

A well-known stochastic geometry framework for wireless cellular networks can be found in~\cite{Andrews2011}; by assuming a
single slope path loss model, the authors have observed the independence of the Signal-to-Interference-plus-Noise-Ratio (SINR) and Spectral Efficiency (SE) from the BS deployment density, where the main consequence is the linear dependence of the ASE on the deployment density.

In some recent work, the effect of cell densification while assuming different models than single slope has been considered. For instance, in~\cite{Galiotto2014} it was shown that by using a combined Line-Of-Sight (LOS)/Non-Line-Of-Sight (NLOS) path loss model the ASE becomes superlinear at low densities and sublinear at high densities. However that work was based on a simulation study, lacking an analytical framework to back the reached conclusions.

In~\cite{Ling2012}, the authors have studied the non-linear behaviour of the ASE with the deployment density in indoor scenarios, and have included an exponential component in the path loss to account for the wall attenuation. As a result, the ASE was shown to scale as $\sqrt{N}$ with the number $N$ of cells. 

In~\cite{Zhang2014}, the authors extended the stochastic geometry framework proposed in~\cite{Andrews2011} to a multi-slope path loss model. They focused their analysis on a dual-slope propagation model that considers two propagation regimes, namely the near-field and the far-field, with different attenuation exponents. The proposed model is particularly suitable for millimeter wave communications, but is essentially different from the one proposed by 3GPP~\cite{3GPP36814} for the assessment of heterogeneous networks in lower frequency bands (e.g., 2 GHz). 

Finally, the effect of NLOS propagation on the outage probability has been studied in~\cite{Bai2014}, where the authors propose a function that gives the probability to have LOS at a given point depending on the distance from the source, on the average size of the buildings, and on the density of the buildings per area. Although it allowed the authors to show the dependence of the outage probability on the density of base stations and of the density of buildings per area,  the propagation model proposed in~\cite{Bai2014} lacks a validation based on a measurement campaign with real data.


\subsection{Our Contribution} 
\label{sub:our_contribution}

In this paper, we focus on a dense network of small cells where the signal propagation considers a path loss model with LOS and NLOS components, and we use stochastic geometry tools to evaluate common performance metrics (such as coverage/outage probability, SE and ASE) as a function of the base station deployment density. Specifically, we consider the propagation model provided by 3GPP for heterogeneous networks at 2  GHz, as we believe this is a reliable model for the performance evaluation of small-cells and dense networks. Our contribution is threefold: (i) we propose a tractable model for the LOS probability function which well fits the 3GPP model and, at the same time, it enables the study of the performance depending on the LOS likelihood of the environment; (ii) we derive a stochastic geometry framework that incorporates the proposed LOS probability function; finally (iii) we study the optimal base station deployment density in terms of the common performance metrics according to the LOS likelihood or,  equivalently, the sparsity of the propagation environment.\footnote{In this paper sparsity refers to the probability of an object being present, in other words the density of objects obstructing the signal propagation. Sparse environments are associate with high probability of LOS.} In particular, we show that in sparser environments characterized by high probability of LOS, the optimal base station density in terms of the lowest outage and the highest spectral efficiency is achieved at lower values of base station density and vice-versa.


\subsection{Paper Structure} 
\label{sub:paper_structure}

The remainder of this paper is organized as follows. In Section~\ref{sect:SystemModel} we describe the system model. We show our formulation for computing the SINR, SE and ASE in Section~\ref{sect:SINR_SE_ASE}. In Section~\ref{sect:results} we present and discuss the results while the conclusions are drawn in Section \ref{sect:conclusions}.

 
\section{System model}\label{sect:SystemModel}

In this paper we consider a network of small-cell base stations deployed
according to a homogeneous and isotropic Spatial Poisson Point Process (SPPP), denoted as $\Phi\subset \mathbb{R}^2$, and of intensity $\lambda$. Further, we assume that each base station (BS) transmits with the same power  $P_{\mathrm{TX}}$, and we focus our analysis on the downlink. Finally, we assume the users to be deployed uniformly over the considered area and that each BS serves at least one user and has full buffer traffic.

\subsection{Channel model} \label{subsect:ChannelModel}

In our analysis, we considered the following path loss model~\cite[Table A.2.1.1.2-3]{3GPP36814} :
\begin{equation}
	\mathrm{PL}(d)=\begin{cases}
		K_{\mathrm{L}}d^{-\beta_{\mathrm{L}}} & \text{with probability}\; p_{\mathrm{L}}(d),\\
		K_{\mathrm{NL}}d^{-\beta_{\mathrm{NL}}} & \text{with probability}\:1-p_{\mathrm{L}}(d),
	\end{cases}\label{eq:propag_outdoor}
\end{equation}
where $\beta_{\mathrm{L}}$ and $\beta_{\mathrm{NL}}$ are the path-loss exponents for LOS and NLOS propagation, respectively; $K_{\mathrm{L}}$ and $K_{\mathrm{NL}}$ are the signal attenuations at distance $d=1$ for LOS and NLOS propagation,\footnote{The parameters $K_{\mathrm{L}}$ and $K_{\mathrm{NL}}$ can either refer to the signal attenuations at distance $d=1$ m or $d=1$ km; this depends on the actual values given for the parameters of the channel model.} respectively; $p_{\mathrm{L}}(d)$ is the probability of having LOS as a function of the distance $d$. We further assume that the propagation is affected by Rayleigh fading, which is exponentially distributed $\sim exp(\mu)$. 

The incorporation of the NLOS component in the path loss model accounts for possible obstructions of the signal due to large scale objects (e.g. buildings), which will result in a higher attenuation of the NLOS propagation compared to the LOS path.


\subsection{LOS probability function}\label{subsect:choosing_p_L}
  
In our study, we refer first to the LOS probability function proposed by 3GPP~\cite[Table A.2.1.1.2-3]{3GPP36814} to assess the network performance in scenarios with pico-cells deployment; this function is the following:
\begin{equation}\label{eq:3GPP_p_L}
	p_{\mathrm{L,3G}}(d) = 0.5-\min\big(0.5,5e^{-\frac{d_0}{d}}\big)+\min\big(0.5, 5e^{-\frac{d}{d_1}}\big).
\end{equation}
Unfortunately, this function would not be practical for an analytical formulation. Therefore, we chose to approximate it with a more tractable one, namely,
\begin{equation}\label{eq:Our_p_L}
	p_{\mathrm{L}}(d) = \exp\left(-(d/L)^2\right)
\end{equation}
where $L$ is a parameter that allows~\eqref{eq:Our_p_L} to be tuned to match~\eqref{eq:3GPP_p_L}, as we will show later on. From a physical stand point, this parameter $L$ can be interpreted as the LOS likelihood of a given propagation environment as a function of the distance. Consider the following motivating example: this could be associated to the density of large scale obstructing objects in the propagation environment. The larger $L$ is, the sparser the environment will be and consequently the higher the probability to have LOS at a given distance from the point of interest and vice-versa. In our study, we will make use of this parameter to analyze the effect of different LOS propagation environments on the network performance.

\section{SINR, spectral efficiency and ASE}\label{sect:SINR_SE_ASE}

In this section we propose an analytical model to compute the SINR Cumulative Distribution Function (CDF) and the ASE of a network where the path-loss includes both Line-of-Sight (LOS) and Non-Line-of-Sight (NLOS) components.

\subsection{Procedure to compute the SINR CCDF}

In order to compute the SINR Complementary CDF (CCDF), we extend the analytical framework first proposed in~\cite{Andrews2011} to include the LOS and NLOS components. From the Slivnyak's Theorem~\cite[Theorem 8.10]{Haenggi2013}, we consider the~\emph{typical user} as the focus of our analysis, which for convenience is assumed to be located at the origin. The procedure is composed of two steps: (i) we compute the SINR CCDF for the typical user conditioned on the distance from the user to the serving base station, denoted as $r$; (ii) using the PDF $f_{r}(R)$ of the distance from the user to the serving BS, we can average the SINR CCDF over all possible values of distance $r$.

Let us denote the SINR by $\gamma$; formally, the CCDF of $\gamma$ is computed as:
\begin{align}\label{eq:SINR_general_def}
	\mathrm{P}\left[\gamma>y\right]&=\mathrm{E_{r}}\big[\mathrm{P}\left[\gamma>y|r\right]\big] \\ \nonumber
	&= \int _{0}^{+\infty}\mathrm{P}\left[\gamma>y|r=R\right]f_{r}(R)\mathrm{d}R.
\end{align}
The key elements of this procedure are the PDF of the distance to the serving base station $f_r(R)$ and the tail probability of the SINR conditioned on $r$, $\mathrm{P}\left[\gamma>y|r=R\right]$. The methodology to compute each of these elements while modelling the LOS and NLOS path loss components will be exposed next.

\subsection{Dual SPPP for the LOS/NLOS propagation model}

The set of the base stations locations originates an SPPP, which we denote by $\Phi=\{x_n\}$.\footnote{Whenever there is no chance of confusion, we drop the subscript $n$ and use $x$ and instead of $x_n$ for convenience of notation.} We recall that the focus of our analysis is the typical user located at the origin. As a result of the propagation model we have adopted in our analysis (see Section~\ref{subsect:ChannelModel}), the user can either be in LOS or NLOS with any base station $x_n$ of $\Phi$. 
Now, we perform the following mapping: we first define the set of LOS points, namely $\Phi_{\mathrm{L}}$, and the set of NLOS points, $\Phi_{\mathrm{NL}}$. Then, each point $x_n$ of $\Phi$ is mapped into $\Phi_{\mathrm{L}}$ if the base station at location $x_n$ is in LOS with the user, while it is mapped into $\Phi_{\mathrm{NL}}$ if the base station at location $x_n$ is in NLOS with the user. Since the probability that $x_n$ is in LOS with the user is  $p_{\mathrm{L}}(\|x\|)$, it follows that each point $x_n$ of $\Phi$ is mapped with probability $p_{\mathrm{L}}(\|x\|)$ into $\Phi_{\mathrm{L}}$ and probability $p_{\mathrm{NL}}(\|x\|) = 1-p_{\mathrm{L}}(\|x\|)$ into $\Phi_{\mathrm{NL}}$.
Given that this mapping is performed independently for each point in $\Phi$, then from the "Thinning Theorem"~\cite[Theorem 2.36]{Haenggi2013} it follows that the processes $\Phi_{\mathrm{L}}$ and $\Phi_{\mathrm{NL}}$ are inhomogeneous SPPPs with density $\lambda_{\mathrm{L}}(x)=\lambda p_{\mathrm{L}}(\|x\|)$
and $\lambda_{\mathrm{NL}}(x)=\lambda\left(1-p_{\mathrm{L}}(\|x\|)\right)$,
respectively.

\subsection{Mapping the NLOS SPPP into an equivalent LOS SPPP} \label{subsect:Mapping}

Given that we have two inhomogeneous SPPP processes, it is not trivial to obtain the distribution of the distance from the user to the serving base station, which will be necessary later on to compute the SINR CDF.
In fact, assuming the user be in LOS with the serving base station at  distance $d_{\mathrm{1}}$, there might be an interfering BS at distance $d_\mathrm{2}<d_{\mathrm{1}}$ which is in NLOS with the user. This is possible because the NLOS propagation is affected by a higher attenuation than the LOS propagation.

Hence, to make our problem more tractable, we map the set of points of the process $\Phi_{\mathrm{NL}}$, which corresponds to the NLOS base stations, into an equivalent LOS process $\Phi_{\mathrm{eq}}$; each point $x\in\Phi_{\mathrm{NL}}$ located at distance $d_{\mathrm{NL}}$ from the user is mapped into a point $x_{\mathrm{eq}}$ located
at distance $d_{\mathrm{eq}}$ from the user, so that the BS located at $x_{\mathrm{eq}}$ provides the same signal power to the user with path-loss $K_{\mathrm{L}}d_{\mathrm{eq}}^{-\beta_{\mathrm{L}}}$ as if the base station were located at $x$ with path-loss $K_{\mathrm{NL}}d_{\mathrm{NL}}^{-\beta_{\mathrm{NL}}}$.

We define the mapping $f_{\mathrm{eq}}$ from  $\Phi_{\mathrm{NL}}$ to $\Phi_{\mathrm{eq}}$ as follows:
\begin{equation}\label{eq:directMapping}
	f_{\mathrm{eq}}:\Phi_{\mathrm{NL}}\rightarrow\Phi_{\mathrm{eq}}\qquad f_{\mathrm{eq}}(x)=\frac{x}{\|x\|}d_{\mathrm{eq}}\left(\|x\|\right),
\end{equation}
\begin{equation}\label{eq:d_eq}
	d_{\mathrm{eq}}(d)=\left(\frac{K_{\mathrm{L}}}{K_{\mathrm{NL}}}\right)^{1/\beta_{\mathrm{L}}}d^{\beta_{\mathrm{NL}}/\beta_{\mathrm{L}}}.
\end{equation}

The inverse function $g_{\mathrm{eq}}$ mapping the points of $\Phi_{\mathrm{eq}}$ to
$\Phi_{\mathrm{NL}}$ is defined as follows:
\begin{equation}\label{eq:inverseMapping}
g_{\mathrm{eq}}=f_{\mathrm{eq}}^{-1}:\Phi_{\mathrm{eq}}\rightarrow\Phi_{\mathrm{NL}}\qquad	g_{\mathrm{eq}}(x)=\frac{x}{\|x\|}d_{\mathrm{eq}}^{-1}\left(\|x\|\right),
\end{equation}%
\begin{equation}\label{eq:inverse_d_eq}
	d_{\mathrm{eq}}^{-1}(d)=\left(\frac{K_{\mathrm{NL}}}{K_{\mathrm{L}}}\right)^{1/\beta_{\mathrm{NL}}}d^{\beta_{\mathrm{L}}/\beta_{\mathrm{NL}}}=K_{\mathrm{eq}}d^{\beta_{\mathrm{eq}}},
\end{equation}
where $K_{\mathrm{eq}}=\left(\frac{K_{\mathrm{NL}}}{K_{\mathrm{L}}}\right)^{1/\beta_{\mathrm{NL}}}$
while $\beta_{\mathrm{eq}}=\beta_{\mathrm{L}}/\beta_{\mathrm{NL}}$. It is important to notice that, for the "Mapping Theorem"~\cite[Theorem 2.34]{Haenggi2013}, $\Phi_{\mathrm{eq}}$ is still an inhomogeneous SPPP.
\vspace{-1mm}
\subsection{PDF of the distance from the user to the serving BS}\label{subsect:PDF_of_distance}

Using the mapping we introduced in Section \ref{subsect:Mapping}, we can
compute the PDF $f_{r}(R)$ of the distance $r$ between the user and the serving BS. We first compute the probability $\mathrm{P}\left[r>R\right]$; the PDF can be ultimately obtained from the derivative of $\mathrm{P}\left[r>R\right]$ as $f_{r}(R)=\frac{\mathrm{d}}{\mathrm{d}R}(1-\mathrm{P}\left[r>R\right])
$.

Let $B(0,l)$ be the ball of radius $l$ centred at the origin $(0,0)$. Moreover, we use the notation $\Phi(\mathcal{A})$ to refer to number of points $x\in \Phi$ contained in $\mathcal{A}$~\cite{Haenggi2013}. Using the mapping we introduced in Section \ref{subsect:Mapping} the probability $\mathrm{P}\left[r>R\right]$ can be found as:
\vspace{-2mm}
\begin{equation*}
	\mathrm{P}\left[r>R\right]=\mathrm{P}\left[ \Phi_{\mathrm{L}}\left(B(0,R)\right)=0\cap\Phi_{\mathrm{eq}}\left(B\left(0,R\right)\right)=0\right]
\end{equation*}
\vspace{-4mm}
\begin{equation*}
	\overset{1}{=}\mathrm{P}\left[\Phi_{\mathrm{L}}\left(B(0,R)\right)=0\cap\Phi_{\mathrm{NL}}\left(B\left(0,d_{\mathrm{eq}}^{-1}(R)\right)\right)=0\right]
\end{equation*}
\vspace{-4mm}
\begin{equation}
	\overset{2}{=}\mathrm{P}\left[\Phi_{\mathrm{L}}\left(B(0,R)\right)=0\right]\cdot\mathrm{P}\left[\Phi_{\mathrm{NL}}\left(B\left(0,d_{\mathrm{eq}}^{-1}(R)\right)\right)=0\right],
\end{equation}
where equality $1$ comes from the mapping defined in \eqref{eq:inverseMapping} and in \eqref{eq:inverse_d_eq}, while equality $2$ comes from the independence of the processes $\Phi_{\mathrm{L}}$ and $\Phi_{\mathrm{NL}}$. By applying the probability function of inhomogeneous SPPP~\cite[Definition 2.10]{Haenggi2013}, we obtain the following,
\vspace{-2mm}
\begin{equation*}
\mathrm{P}\left[r>R\right]=\exp\bigg(-\int_{B(0,R)}\lambda_{\mathrm{L}}(x)\mathrm{d}x\bigg)
\end{equation*}
\vspace{-3mm}
\begin{equation}\label{eq:prob_r_greater_than_R}	\exp\bigg(-\int_{B\left(0,d_{\mathrm{eq}}^{-1}(R)\right)}\lambda_{\mathrm{NL}}(x)\mathrm{d}x\bigg).
\end{equation}
%
From \eqref{eq:prob_r_greater_than_R}, we can obtain $f_{r}(R)$, first, by integrating and, second, by computing its first derivative with respect to $R$. The formulation in~\eqref{eq:prob_r_greater_than_R} is general and thus can be applied to several LOS probability functions $p_{\mathrm{L}}(d)$. As we introduced earlier in Section \ref{subsect:choosing_p_L}, in this paper we consider use \eqref{eq:Our_p_L} as LOS probability function. The details and the final expression for $f_{r}(R)$ are provided in the Appendix.

\subsection{SINR complementary cumulative distribution function}\label{subsect:SINR_ICDF_Final}

The probability $\mathrm{P}\left[\gamma>y|r=R\right]$ can be computed as in~\cite[Theorem 1]{Andrews2011}; we skip the details and provide the general formulation:
\vspace{-2mm}
\begin{align}\label{eq:P_SINR_geq_y_conditioned}
	\mathrm{P}\left[\gamma>y|r=R\right]&=\mathrm{P}\left[\frac{g K_{\mathrm{L}}R^{-\beta_{\mathrm{L}}}}{\sigma^{2}+I_{R}}>y\right] \\ \nonumber
	&=e^{-\mu y K_{\mathrm{L}}^{-1}R^{\beta_{\mathrm{L}}}\sigma^{2}}\mathcal{L}_{I_{R}}(\mu y K_{\mathrm{L}}^{-1}R^{\beta_{\mathrm{L}}}),
\end{align}
where $g$ is the Rayleigh fading, which we assume to be an exponential random variable $\sim exp(\mu)$; $\sigma^{2}$ is the variance of the additive white Gaussian noise normalized with the respect to the transmit power;  $I_{R}$ is the interference conditioned on $R$, i.e.,
%
%
%
$I_{R} = \sum\limits _{\{i:\: x_i\in\Phi_{\mathrm{L},R}\}}g_i K_{\mathrm{L}}\|x_i\|^{-\beta_{\mathrm{L}}} + \sum\limits _{\{j:\: x_j\in\Phi_{\mathrm{eq},R}\}}g_j K_{\mathrm{L}}R_j^{-\beta_{\mathrm{L}}}$,
where $g_i$ and $g_j$ are independent and identically distributed $\sim \exp(\mu)$ fading coefficients;  $\Phi_{\mathrm{L},R}=\{x\in\Phi_{\mathrm{L}} : \|x\|>R\}$; $\Phi_{\mathrm{eq},R}=\{x\in\Phi_{\mathrm{eq}} : \|x\|>R\}$.   
Hence, compared to the formulation of $\mathcal{L}_{I_{R}}(s)$ proposed in \cite{Andrews2011}, in our case we have to deal with two inhomogeneous SPPP, namely $\Phi_{\mathrm{L}}$ and $\Phi_{\mathrm{NL}}$ (or the equivalent $\Phi_{\mathrm{eq}}$) instead of with a single homogeneous SPPP. 
The Laplace transform $\mathcal{L}_{I_{R}}(s)$ can be written
as follows:
\vspace{-1mm}	
\begin{equation*}
\mathcal{L}_{I_{R}}(s)
	=\mathrm{E}_{\Phi_{\mathrm{L}},\Phi_{\mathrm{eq}},g_i,g_j}\bigg[\exp\bigg(-s\sum\limits _{\{i:\: x_i\in\Phi_{\mathrm{L},R}\}}g_i K_{\mathrm{L}}\|x_i\|^{-\beta_{\mathrm{L}}}\bigg)
\end{equation*}
\vspace{-2mm}	
\begin{equation*}	
	\exp\bigg(-s\sum\limits _{\{j:\: x_j\in\Phi_{\mathrm{eq},R}\}}g_j K_{\mathrm{L}}\|x_j\|^{-\beta_{\mathrm{L}}}\bigg)\bigg].
\end{equation*}
%
%
By applying the mapping defined in Section \ref{subsect:Mapping} and because $\Phi_{\mathrm{L}}$ is independent of $\Phi_{\mathrm{NL}}$, we obtain:
\vspace{-1mm}
\begin{equation*}
\mathcal{L}_{I_{R}}(s)=\mathrm{E}_{\Phi_{\mathrm{L}},g_i}\bigg[\exp\bigg(-s\sum\limits _{\{i:\: x_i\in\Phi_{\mathrm{L},R}\}}g_i K_{\mathrm{L}}\|x_i\|^{-\beta_{\mathrm{L}}}\bigg)\bigg] 
\end{equation*}
\vspace{-2mm}
\begin{equation*}
\mathrm{E}_{\Phi_{\mathrm{NL}},g_j}\bigg[\exp\bigg(-s\sum\limits _{\{j:\: x_j\in\Phi_{\mathrm{NL}},\: \|x_j\|>d_{\mathrm{eq}}^{-1}(R)\}}g_j K_{\mathrm{NL}}\|x_j\|^{-\beta_{\mathrm{NL}}}\bigg)\bigg].
\end{equation*}
%
The Probability Generating Functional (PGFL) for SPPP holds
also in case of inhomogeneous SPPP; therefore, we obtain:
\vspace{-2mm}
\begin{equation*}
\mathcal{L}_{I_{R}}(s)=\exp\Bigg(-2\pi\lambda\int\limits _{R}^{+\infty}\left[\frac{sK_{\mathrm{L}}v^{-\beta_{\mathrm{L}}}}{sK_{\mathrm{L}}v^{-\beta_{\mathrm{L}}}+\mu}\right]p_{\mathrm{L}}(v)v\mathrm{d}v\Bigg) 
\end{equation*}
\vspace{-4mm}
\begin{equation}\label{eq:laplace_solved}
\exp\Bigg(-2\pi\lambda\int\limits _{d_{\mathrm{eq}}^{-1}(R)}^{+\infty}\left[\frac{sK_{\mathrm{NL}}v^{-\beta_{\mathrm{NL}}}}{sK_{\mathrm{NL}}v^{-\beta_{\mathrm{NL}}}+\mu}\right]p_{\mathrm{NL}}(v)v\mathrm{d}v\Bigg).
\end{equation}
Eq. \eqref{eq:laplace_solved} along with \eqref{eq:prob_r_greater_than_R} and \eqref{eq:f_R_final} can be plugged in \eqref{eq:SINR_general_def} to obtain the SINR CCDF through numerical integration.

\subsection{Average Spectral Efficiency and Area Spectral Efficiency}

Similarly to~\cite[Section IV]{Andrews2011} we compute the average spectral efficiency and the ASE of the network. First, we define the ASE as:
\begin{equation}\label{eq:ASE}
	\mathrm{ASE}=\frac{\#\mathrm{BS}\cdot\mathrm{BW}\cdot\overline{\mathrm{C}}}{A\cdot\mathrm{BW}}=\frac{\lambda\cdot A\cdot\mathrm{BW}\cdot\overline{\mathrm{C}}}{A\cdot\mathrm{BW}}=\lambda\cdot\overline{\mathrm{C}},
\end{equation}
where $\mathrm{BW}$ is the bandwidth, $\mathrm{\overline{C}}$ is the average spectral efficiency, $A$ is the area, $\#\mathrm{BS}$ is the number of base stations within the area $A$. 

The average rate $\mathrm{\overline{C}}$ can be computed as follows \cite{Andrews2011}:
\begin{align}
	\mathrm{\overline{C}}=\mathrm{E}\left[\log_{2}(1+\gamma)\right]=\int _{0}^{+\infty}\mathrm{P}\left[\log_{2}(1+\gamma)>u\right]\mathrm{d}u \\ \nonumber
	=\int_{0}^{+\infty}\int _{0}^{+\infty}\mathrm{P}\left[\log_{2}(1+\gamma)>u|r=R\right]f_{r}(R)\mathrm{d}R\mathrm{d}u.
\end{align}
If we develop $\mathrm{\overline{C}}$ further, we obtain:%
\begin{align}\label{eq:rate_final}
	\mathrm{\overline{C}}=\int _{0}^{+\infty}\int _{0}^{+\infty}e^{-\mu(2^{u}-1)K_{\mathrm{L}}^{-1}R^{\beta_{\mathrm{L}}}\sigma^{2}} \\ \nonumber
	\mathcal{L}_{I_{R}}\big(\mu(2^{u}-1)K_{\mathrm{L}}^{-1}R^{\beta_{\mathrm{L}}}\big)f_{r}(R)\mathrm{d}R\mathrm{d}u
\end{align}
where $\mathcal{L}_{I_{R}}(s)$ is given in \eqref{eq:laplace_solved}. Similarly to the SINR CCDF, \eqref{eq:rate_final} can be evaluated numerically. 

\section{Results}\label{sect:results} 

In this section we present and discuss the results we obtained by integrating numerically the expressions of the SE, of the ASE and of the outage probability. We also compare the semi-analytical results with those obtained through Monte Carlo simulations in order to gauge the fitness of our model. 

In our study we assume the network is interference limited, neglecting the thermal noise. 
For the channel model, we used \eqref{eq:propag_outdoor} and, for the associated parameters, we set $K_{\mathrm{L}}=10^{10.38}$, $\beta_{\mathrm{L}}=2.09$, $K_{\mathrm{NL}}=10^{14.54}$, $\beta_{\mathrm{NL}}=3.75$, $d_0 =156\mathrm{m}$, and $d_1 = 30\mathrm{m}$ as per the 3GPP model for evaluation of pico-cellular networks~\cite[Table A.2.1.1.2-3]{3GPP36814}. 

\subsection{Validation of the LOS probability function}\label{subsect:validation_of_p_L}

\begin{figure}
\centering
\includegraphics[width=.89\columnwidth]{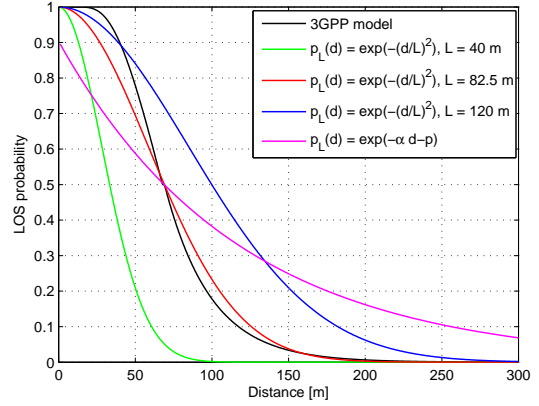}
\caption{The function $p_{\mathrm{L}}(d) = exp(-\alpha d - p)$ \cite{Bai2014} has been plotted to intersect the 3GPP model at the point corresponding to probability 0.5. The corresponding values $\alpha$ and $p$ are $8.59\cdot 10^{-3}$ and $1.01\cdot 10^{-1}$, respectively.}\label{fig:p_LOS_functions}
\vspace{-3mm}
\end{figure}

In this subsection we explain the procedure to validate the LOS probability function we adopted in our model and we also provide the results of this validation. We first show the two LOS probability functions, namely~\eqref{eq:Our_p_L} and~\eqref{eq:3GPP_p_L}, in Fig.~\ref{fig:p_LOS_functions}. To match the 3GPP model, we set $L = 82.5$ m so that \eqref{eq:3GPP_p_L} and~\eqref{eq:Our_p_L} intersect at the point corresponding to probability 0.5. In Fig. \ref{fig:p_LOS_functions},~\eqref{eq:Our_p_L} is also shown for other values of the parameter $L$, i.e. $L=120$ m and $L=40$ m. Moreover, we also plotted the LOS probability function,
\begin{equation}\label{eq:Bai2014}
	p_{\mathrm{L}}(d)=e^{-\left( \alpha d - p \right)}
\end{equation}
proposed by \cite{Bai2014} to check whether this function provides results that are with in line with those obtained with the 3GPP model.

\begin{figure}
	\centering
	\includegraphics[width=.89\columnwidth]{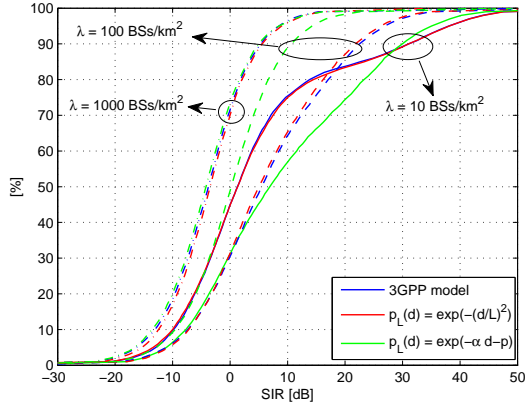}
	\caption{Comparison of SIR CDFs for different functions $p_{\mathrm{L}}(d)$.}\label{fig:SIR_CDFs}
\vspace{-5mm}	
\end{figure}

The validation of~\eqref{eq:Our_p_L} is only carried out for $L = 82.5$ m, whereas the other two values of the parameter $L$ will solely be used to study the effect of different propagation environments on the network performance. To benchmark the LOS probability function we compute the Signal-to-Interference-Ratio (SIR) CDF by numerical integration of \eqref{eq:SINR_general_def}, while assuming~\eqref{eq:Our_p_L} as $p_{\mathrm{L}}(d)$; we then compare this curve with the SIR CDF evaluated through Monte Carlo simulations assuming \eqref{eq:3GPP_p_L} as $p_{\mathrm{L}}(d)$. The benchmark is shown in Fig.~\ref{fig:SIR_CDFs}.
As we can from this plot, the two CDFs well match for different values of BS densities; this implies that our model is well suited to study the effect of the LOS/NLOS propagation on the performance of dense networks, as it yields results in line with those obtained with the 3GPP model. 
On the contrary, the SIR obtained with~\eqref{eq:Bai2014} as $p_{\mathrm{L}}(d)$ only matches the 3GPP benchmark for high cell densities; in fact, at high densities the strongest interferers  are in LOS with the user regardless of the shape of the $p_{\mathrm{L}}(d)$ functions we have tested. Using \eqref{eq:Bai2014} as probability LOS function would give us accurate results only for a limited range of BSs density and, therefore, would not be suitable for the kind of study we carry out in our paper.

\subsection{Spectral efficiency, ASE and outage probability}\label{subsect:rate_ASE_density}

To study the performance of the dense networks as a function of the BS density, we evaluated numerically the SE (eq. \eqref{eq:rate_final}), the ASE (eq.\eqref{eq:ASE}) and the outage probability (obtained from eq. \eqref{eq:SINR_general_def}).

\begin{figure}
	\centering
	\includegraphics[width=.89\columnwidth]{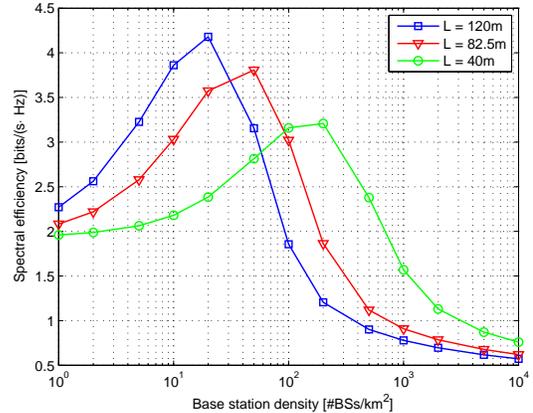}
	\caption{Spectral efficiency vs base station density.}\label{fig:Rate_vs_density}
	\vspace{-3mm}
\end{figure}

We first present the SE results, which are shown in Fig. \ref{fig:Rate_vs_density}.
From this plot, we can see that the SE curve is not constant but exhibits a peak for a given value of base station density. This result is in contrast with what we would obtain if we used a single slope propagation model, for which the SE has been shown to be constant and thus independent of the base station density~\cite{Andrews2011}. Therefore, we can infer that the reason for observing a non constant behaviour of the SE curve as a function of the BS density is the LOS/NLOS propagation. 

The peak of the curve is observed for a given value of the BS density which we will refer to as \textit{optimal BS density for the SE}, denoted by $\lambda^*_{\mathrm{SE}}$. At the optimal BS density, the user is likely to be in LOS with the serving BS, while it is in NLOS with most of the interferers, meaning that the interference power is low. For BS densities lower than $\lambda^*_{\mathrm{SE}}$, the serving BS is likely to be in NLOS with the users with a consequent reduction of the received power, of the SINR and therefore of the SE. On the contrary, at BS densities higher than $\lambda^*_{\mathrm{SE}}$, the interfering BSs are likely to be in LOS with the user, causing an overall interference growth and thus a reduction of the SIR and SE. 

The behaviour of the SE curve is also influenced by the LOS probability as a function of the distance. In fact, looking at the plots in Fig.~\ref{fig:Rate_vs_density}, we notice that the optimal SE density depends on the LOS likelihood parameter L. In sparser propagation environments (e.g., $L = 120$m) the propagation is likely to be of the LOS kind at longer distances from the user, compared to the case of dense propagation environments; this means that $\lambda^*_{\mathrm{SE}}$ will be reached at a lower BS density.

\begin{figure}
	\centering
	\includegraphics[width=.89\columnwidth]{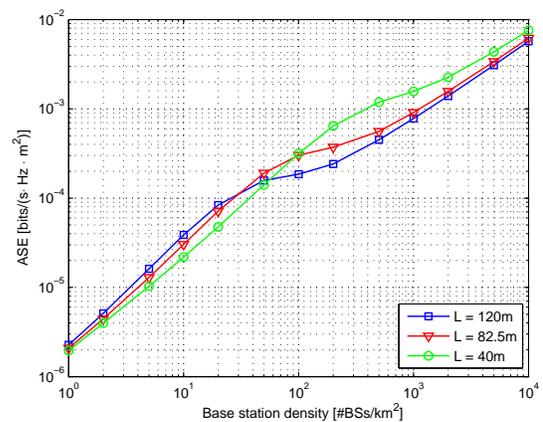}
	\caption{ASE vs base station density.}\label{fig:fig1}
	\vspace{-5mm}
\end{figure}

\begin{figure}
	\centering	\includegraphics[width=.89\columnwidth]{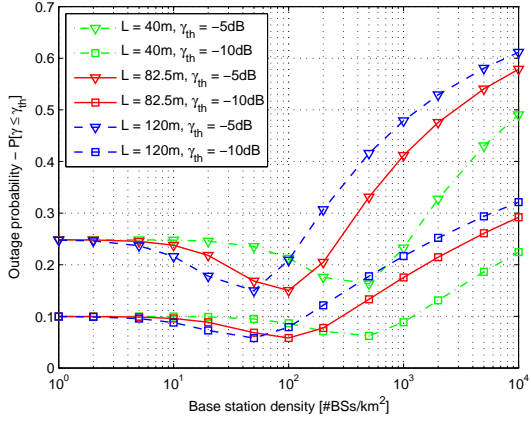}
	\caption{Outage probability vs base station density.}\label{fig:fig3}
	\vspace{-5mm}
\end{figure}

Fig.~\ref{fig:fig1} shows the ASE curves for different LOS propagation environments.
Unlike the SE, the ASE is an increasing function of the BS density $\lambda$ and this is due to the effect of the cell densification, as shown in~\eqref{eq:ASE}. Nonetheless, the SE has an impact on the ASE, as we can notice from the lower steepness of the ASE curve. In particular, the ASE switches from a superlinear gain at low cell densities to a sublinear gain at high cell-densities. 
To complete the analysis of the performance of cell densification, we also need to evaluate the outage probability curves, shown in Fig. \ref{fig:fig3}.
From this plot we can see that, at higher BS densities, the outage starts growing drastically and, depending on the LOS likelihood, it can reach 22-32\% for -10dB SIR threshold and 50-60\% for -5dB SIR threshold. This is due to the effect of the interfering BSs which, at high cell densities, are likely to be in LOS with the user. This will be perceived by the user as a strong interference, which causes an SIR reduction. In addition, let us notice that in environments with higher NLOS probability, the optimum of the outage curve is achieved at higher BS densities.

In the light of our study, we make the argument that, despite the high gain provided in terms of cell ASE, cell densification can severely affect the outage probability and hence the network coverage. Therefore, if ones needs an extremely dense cell deployment to meet some given capacity requirements, networks will require interference handling techniques in order to operate in a satisfactory manner. 

\section{Conclusions}\label{sect:conclusions}

In this paper we have proposed a stochastic geometry framework to study the coverage/outage probability, the Spectral Efficiency (SE) and the Area Spectral Efficiency (ASE) of dense small cell networks where the signal path-loss model includes both LOS and NLOS components. Through our formulation, we have investigated how the network performance scales with the density of the base station deployment and we studied the effect of the propagation environment in terms of LOS likelihood on SE, ASE and outage probability. 

Finally, we have shown that propagation environments with high NLOS probability allow for a denser cell deployment. Overall, in this paper we sought to uncover insights on how dense networks should be deployed while taking into account a model that more accurately describes the signal propagation compared to the more common single slope path-loss, which is the current standard reference model when studying wireless networks.

%

\appendices{
\numberwithin{equation}{section}

\section{PDF of the distance to the serving base station}

If we assume the LOS probability to be given by~\eqref{eq:Our_p_L},
we can further develop \eqref{eq:prob_r_greater_than_R} as follows:
$$
\mathrm{P}\left[r>R\right]=\exp\bigg(-\lambda\int_{B(0,R)}p_{\mathrm{L}}(\|x\|)\mathrm{d}x\bigg)$$
\begin{equation}\label{eq:Prob_r_gt_R_integral}\exp\bigg(-\lambda\int_{B\left(0,d_{\mathrm{eq}}^{-1}(R)\right)}\left(1-p_{\mathrm{L}}(\|x\|)\right)\mathrm{d}x\bigg).
\end{equation}
By solving the integrals in \eqref{eq:Prob_r_gt_R_integral} and with further symbolic manipulation we obtain:
\begin{equation}\mathrm{P}\left[r>R\right]=e^{\pi\lambda L^{2}e^{-\frac{R^{2}}{L^{2}}}}\cdot e^{-\pi\lambda L^{2}e^{-\frac{R_{\mathrm{eq}}^{2}}{L^{2}}}}\cdot e^{-\pi\lambda R_{\mathrm{eq}}^{2}},
\end{equation}
where $R_{\mathrm{eq}} = d_{\mathrm{eq}}^{-1}(R)$. Let us define the functions $f_{1}(R)$, $f_{2}(R)$, $f_{3}(R)$ and their first derivatives  $f_{1}^{\prime}(R)$,
$f_{2}^{\prime}(R)$, and $f_{3}^{\prime}(R)$, respectively, as follows: 
$$
f_{1}(R)=e^{\pi\lambda L^{2}e^{-\frac{R^{2}}{L^{2}}}},\qquad f_{2}(R)=e^{-\pi\lambda L^{2}e^{-\frac{R_{\mathrm{eq}}^{2}}{L^{2}}}},
$$
$$
f_{3}(R)=e^{-\pi\lambda R_{\mathrm{eq}}^{2}}, \qquad f_{1}^{\prime}(R)=-2\pi\lambda Re^{-\frac{R^{2}}{L^{2}}}f_{1}(R),
$$
$$
f_{2}^{\prime}(R)=\pi\lambda K_{\mathrm{eq}}^{2}2\beta_{\mathrm{eq}}R^{2\beta_{\mathrm{eq}}-1}e^{-\frac{-K_{e\mathrm{q}}^{2}R^{2\beta_{\mathrm{eq}}}}{L^{2}}}f_{2}(R),
$$
$$
f_{3}^{\prime}(R)=-\pi\lambda K_{\mathrm{eq}}^{2}2\beta_{\mathrm{eq}}R^{2\beta_{\mathrm{eq}}-1}f_{3}(R).
$$
Then we can write $\mathrm{P}\left[r>R\right]$ as $\mathrm{P}\left[r>R\right] = f_{1}(R)f_{2}(R)f_{3}(R)$. The PDF of the distance from the user to the serving base station is given by:
$$
f_{r}(R)=-\frac{\mathrm{d}}{\mathrm{d}R}\left(\mathrm{P}\left[r>R\right]\right)=-\frac{\mathrm{d}}{\mathrm{d}R}\left[f_{1}(R)f_{2}(R)f_{3}(R)\right]
$$
\begin{equation*}
=-\big(f_{1}^{\prime}(R)f_{2}(R)f_{3}(R)+f_{1}(R)f_{2}^{\prime}(R)f_{3}(R)
\end{equation*}
\begin{equation}\label{eq:f_R_final}
+f_{1}(R)f_{2}(R)f_{3}^{\prime}(R)\big).
\end{equation}

}

\section*{Acknowledgements}
This work was funded by Higher Education Authority under grant HEA/PRTLI Cycle 5 Strand 2 TGI and by Science Foundation Ireland through CTVR CSET grant number 10/CE/I1853.



\begin{thebibliography}{1}
\providecommand{\url}[1]{#1}
\csname url@samestyle\endcsname
\providecommand{\newblock}{\relax}
\providecommand{\bibinfo}[2]{#2}
\providecommand{\BIBentrySTDinterwordspacing}{\spaceskip=0pt\relax}
\providecommand{\BIBentryALTinterwordstretchfactor}{4}
\providecommand{\BIBentryALTinterwordspacing}{\spaceskip=\fontdimen2\font plus
\BIBentryALTinterwordstretchfactor\fontdimen3\font minus
  \fontdimen4\font\relax}
\providecommand{\BIBforeignlanguage}[2]{{%
\expandafter\ifx\csname l@#1\endcsname\relax
\typeout{** WARNING: IEEEtran.bst: No hyphenation pattern has been}%
\typeout{** loaded for the language `#1'. Using the pattern for}%
\typeout{** the default language instead.}%
\else
\language=\csname l@#1\endcsname
\fi
#2}}
\providecommand{\BIBdecl}{\relax}
\BIBdecl

\bibitem{Andrews2011}
J.~G. Andrews, F.~Baccelli, and R.~K. Ganti, ``{A Tractable Approach to
  Coverage and Rate in Cellular Networks},'' \emph{IEEE Trans. Wireless
  Commun.}, vol.~59, no.~11, pp. 3122--3134, 2011.

\bibitem{Galiotto2014}
C.~Galiotto, I.~M. Gomez, N.~Marchetti, and L.~Doyle, ``{Effect of LOS/NLOS
  Propagation on Area Spectral Efficiency and Energy Efficiency of
  Small-Cells},'' \emph{IEEE Global Communications Conference (GLOBECOM) 2014},
  pp. 3471 -- 3476, 2014.

\bibitem{Ling2012}
J.~Ling and D.~Chizhik, ``{Capacity Scaling of Indoor Pico-Cellular Networks
  via Reuse},'' \emph{IEEE Commun. Lett.}, vol.~16, no.~2, pp. 231--233, Feb.
  2012.

\bibitem{Zhang2014}
\BIBentryALTinterwordspacing
X.~Zhang and J.~G. Andrews, ``{Downlink Cellular Network Analysis with
  Multi-slope Path Loss Models},'' 2014. [Online]. Available:
  \url{http://arxiv.org/abs/1408.0549}
\BIBentrySTDinterwordspacing

\bibitem{Bhushan2014}
N.~Bhushan, L.~Junyi, D.~Malladi, R.~Gilmore, D.~Brenner, A.~Damnjanovic,
  R.~Sukhavasi, and S.~Geirhofer, ``{Network Densification: the Dominant Theme
  for Wireless Evolution into 5G},'' \emph{IEEE Communications Magazine}, 2014.

\bibitem{3GPP36814}
3rd Generation Partnership Project~(3GPP), ``{F}urther {A}dvancements for
  {E}-{UTRA} {P}hysical {L}ayer {A}spects ({R}elease 9),'' Mar. 2010, 3GPP TR
  36.814 V9.0.0 (2010-03).

\bibitem{Bai2014}
T.~Bai, R.~Vaze, and R.~W. Heath, ``{Analysis of Blockage Effects on Urban
  Cellular Networks},'' \emph{IEEE Trans. Wireless Commun.}, vol.~13, no.~9,
  pp. 5070--5083, Sep. 2014.

\bibitem{Haenggi2013}
M.~Haenggi, \emph{Stochastic Geometry for Wireless Networks}.\hskip 1em plus
  0.5em minus 0.4em\relax Cambridge Press, 2013.

\end{thebibliography}
\end{document}